\documentclass[journal=jcisd8,manuscript=article]{achemso}
\SectionNumbersOn
\usepackage[version=3]{mhchem} 
\usepackage{xcolor} 
\usepackage{tikz}
\usepackage{adjustbox}
\usetikzlibrary{shapes,positioning, fit, shadows, backgrounds, calc}

\newcommand*{\tpas}[1]{$\sigma_{#1}$}
\newcommand*{\osc}[1]{$f_{#1}$}
\newcommand*{\crit}[1]{{\bf Criterion #1}}
\newcommand*{\exe}[1]{$\omega_{#1}$}

\author{Alexander Mielke}
\affiliation[University of Regensburg]
{Institute of Chemistry and Pharmacy, University of Regensburg, Universitaetsstrasse 31, 93041 Regensburg, Germany}
\altaffiliation{Contributed equally to this work}
\author{Alexander Scrimgeour}
\affiliation[University of Regensburg]
{Institute of Chemistry and Pharmacy, University of Regensburg, Universitaetsstrasse 31, 93041 Regensburg, Germany}
\altaffiliation{Contributed equally to this work}
\author{Enrico Tapavicza}
\affiliation[University of Regensburg]
{Institute of Chemistry and Pharmacy, University of Regensburg, Universitaetsstrasse 31, 93041 Regensburg, Germany}
\alsoaffiliation{Department of Chemistry and Biochemistry, California State University, Long Beach, 1250 Bellflower Boulevard, Long Beach, California 90840-9507, United States}
\email{enrico.tapavicza@ur.de}

\title[Chemical Space of Molecular Nanomotors]
  {Chemical Space of Molecular Nanomotors: Optimizing Photochemical Properties for One- and Two-photon Applications}

\abbreviations{IR,NMR,UV}

\begin{document}

\begin{tocentry}
\includegraphics{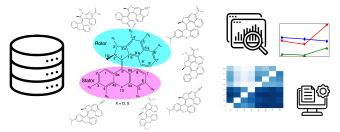}
Some journals require a graphical entry for the Table of Contents.
This should be laid out ``print ready'' so that the sizing of the
text is correct.

Inside the \texttt{tocentry} environment, the font used is Helvetica
8\,pt, as required by \emph{Journal of the American Chemical
Society}.

The surrounding frame is 9\,cm by 3.5\,cm, which is the maximum
permitted for  \emph{Journal of the American Chemical Society}
graphical table of content entries. The box will not resize if the
content is too big: instead it will overflow the edge of the box.

This box and the associated title will always be printed on a
separate page at the end of the document.

\end{tocentry}

\begin{abstract}
Light-driven molecular nanomotors hold promise for applications in material science and biomedicine. Significant efforts have focused on improving their efficiency, often targeting single candidate molecules. Here, we present a systematic data-driven approach to design nanomotors with high isomerization quantum yields for one- and two-photon applications, the latter being critical for biomedical applications requiring near-infrared light. 
We analyze the excited state properties of a dataset of 2016 nanomotors substituted with electron-donating and electron-withdrawing (push-pull) groups. Among the the top candidates, we achieved an increase in two-photon absorption strengths of up to two orders of magnitude compared to existing nanomotors. 
To ensure that the $\pi-\pi^*$-character of the excited state is preserved, which is necessary to achieve the required photoisomerization, we introduce a photoreactivity score, that gauges the excited state character based on the transition. Furthermore, we benchmark three machine learning (ML) models—Kernel Ridge Regression, XGBoost, and a Neural Network—using physical and connectivity-based molecular descriptors. 
The excellent accuracy of our ML predictions holds promise to replace computationally costly quantum chemistry calculations in chemical space explorations.

\end{abstract}

 
\section{Introduction}

\begin{figure}[H]
    \centering
    \includegraphics[width=3.3in]{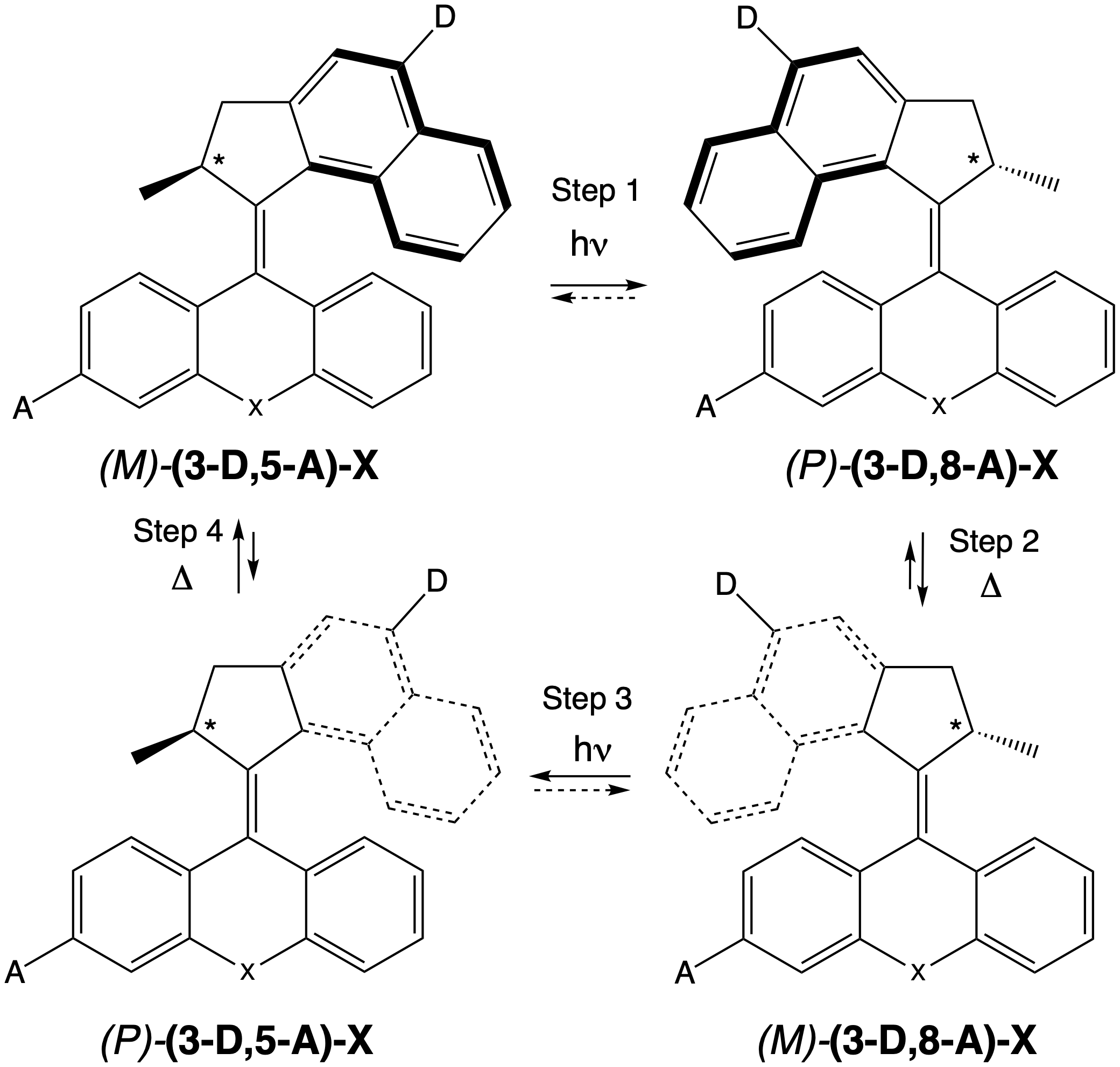}
    \caption{Mechanism of unidirectional rotation of an overcrowded alkene MNM, here exemplified for the generic isomer pair {\bf (3-D,5-A/3-D,8-A)-X}. Photochemical steps are denoted by $h\nu$, thermal steps are denoted by $\Delta$. D and A are unspecified donor and acceptor substituents, respectively. The chiral center is marked with an asterisk.}
    
    \label{fig:scheme1}
\end{figure}

In the last two decades, the development of light-driven 
molecular nanomotors (MNMs) has led to a variety of possible applications, such as mesoscale photoresponsive materials \cite{deng2024photo} and applications in biology \cite{garcia2017molecular} and medicine \cite{garcia2019light}. Similar to photoswitches, which have been extensively proposed for medical applications\cite{velema2014photopharmacology,grathwol2019azologization}, light-driven MNMs also have the advantage to be controllable with high spatial and temporal precision.
A large number of the {\it second generation} MNMs are based on the 
overcrowded alkene architecture\cite{feringa1999,koumura2002second,pooler2021designing} (Figure~\ref{fig:scheme1}). To achieve unidirectional rotation, photochemical EZ-isomerization ($h\nu$) alternates with thermal helix inversion (THI, $\Delta$), as shown in Figure~\ref{fig:scheme1}.

A major objective in designing MNMs is to increase the frequency of rotation, which entails both increasing EZ-photoisomerization quantum yield and THI rates. While increasing the THI rates has been the subject of several studies,\cite{koumura2002second,unidirectionality} here we focus on maximizing the photochemical EZ-isomerization quantum yield. The photochemical isomerization quantum yield depends on two key quantities:\cite{Thompson2018,Tapavicza2018} i) the amount of photons absorbed by the MNM, and ii) the probability of successful EZ-photoisomerization once the molecule is excited. 
In case of a typical one-photon absorption, the oscillator strength \osc{} describes the ability to absorb photons. A prime objective in designing efficient MNMs is therefore increasing the oscillator strength. However, applications in biological tissue often require near infrared (NIR) excitation due to its better penetration depth and lower phototoxiticity.\cite{ash2017effect} 
Recently large progress has been made in developing photoresponsive materials with one-photon excitation energies in the NIR.\cite{hartingertriplet}
The typical energy gap of overcrowded alkenes, however, lies in the UV/Vis region. Therefore, excitation with NIR radiation can only be achieved by a two-photon process, leading to a doubling of the corresponding one-photon excitation wavelength (or halfing of the energy gap). 
Although two-photon processes pose challenges in practical applications due to their low sensitivity, a variety of molecular sensitizers for photodynamic therapy for two-photon absorption (TPA) have been developed \cite{bolze2017molecular}.
Recently several prototype MNMs have been harnessed to function by two-photon excitations.\cite{Liu2019ACSNano,GUINART2024115649}
However, due to the low two-photon absorption strength (TPAS), this process still requires high radiation intensity, which can cause unwanted tissue damage in biological applications.\cite{alabugin2019} Therefore, increasing TPAS is a major route to pave the way for two-photon applications in photodynamic therapy. 

Once the molecule is promoted to the excited state, it still needs to undergo the desired photochemical transformation, to achieve the desired mechanical motion. In the case of EZ-isomerization, the probability of isomerization is related to the orbital character of the electronic transition: a $\pi-\pi^*$ character typically leads to the desired double-bond EZ-isomerization \cite{lucia2025first}.
Thus, increasing absorption strength does not necessarily lead to increased quantum yield. One still needs to ensure that the state reached by excitation has the required $\pi-\pi^\ast$ orbital character.

\begin{figure}[H]
    \centering
    \includegraphics[width=3.3in]{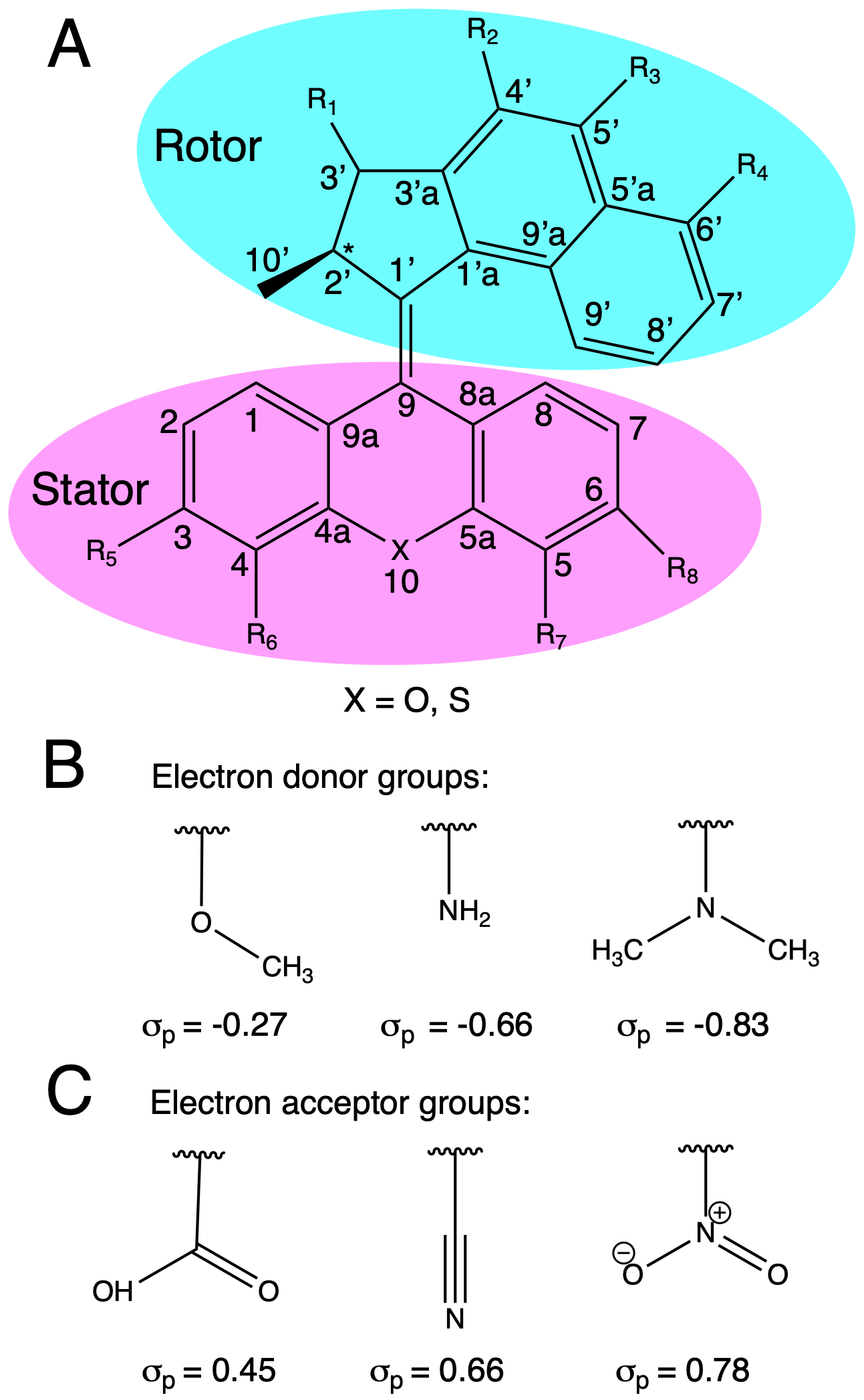}
    \caption{(A) Base structure of the overcrowded alkene MNM with atom numbering. Positions R$_{1}$-R$_{8}$ are used as possible sites for substituents. For each candidate molecule, one position is substituted with one electron donor group (B) and another position is substituted with one electron acceptor group (C). We restrict our study to molecules with one donor-acceptor pair only. Hammett parameters ($\sigma_{p}$), indicating electron-donating or electron-withdrawing strengths are given\cite{hansch1991survey}.}
    \label{ds2g_mol}
\end{figure}

Lately, a large effort has been undertaken in applying computational tools to study the mechanism of MNMs and to increase their efficiency \cite{stauch2016predicting,Nikiforov2016,oruganti2015computational,filatov_rotor,filatov_rotor_2,filatov_rotor_3,leticia_rotor,lucia2025first}. 
In an attempt to initiate a data-driven approach\cite{vela2019exploring,tapavicza2021elucidating} to identify promising candidates for nanomotor applications, we systematically explore the chemical space of substituted MNMs.
Since one of our objectives is to apply MNM to TPA, we study derivatives of a Feringa-type\cite{augulis2009light} overcrowded alkene MNM substituted with electron donor-acceptor pairs (push-pull systems) (Figure~\ref{ds2g_mol}), which have been shown to effectively increase the TPAS\cite{albota1998design,TPADesign2009Pawlicki}.

To this end, we generated a comprehensive molecular dataset of 2016 molecules and analyzed the impact of substituent groups on photochemical excited-state properties (Figure~\ref{ds2g_mol}). 

For all generated structures, excited state quantum mechanical calculations were performed using time-dependent density functional theory (TDDFT)\cite{Runge84}. The effects of the donor and acceptor groups and occupied positions on the excitation energies \exe{}, oscillator strengths \osc{}, and TPAS \tpas{} values were analyzed. An additional analysis of the transition densities was also performed to identify the photochemically active $\pi - \pi^*$ state. We define a set of criteria that a molecule has to fulfill, to identify candidates for potential application in biological tissue.
However, depending on the positions of the substituents, the pair of stable (or metastable) structures involved in the rotation cycle might be chemically different, as suggested by the example in Figure~\ref{fig:scheme1}: isomer {\it (M)-}{\bf (3-B,5-A)-X} (and its conformer {\it (P)}-{\bf (3-B,5-A)-X}) is chemically different from its corresponding isomer {\it (M)}-{\bf (3-B,8-A)-X} (and its conformer {\it (P)}-{\bf (3-B,8-A)-X}). Thus, to ensure proper functioning of the rotational cycle, 
both isomers involved in a particular rotational cycle must fulfill the criteria. In the following, we will refer to a molecule and its corresponding isomer as an {\it isomer pair}.
The defined criteria for the one- or two-photon excitation are: 
\begin{enumerate}
    \item High Photon Harvesting Efficiency: 
 Large \osc{} or \tpas{} for both stable conformers ({\it (M)}-conformers in the case of the absolute configuration used here), ensuring that a large percentage of the irradiated photons are harvested. (\crit{1})
    \item Unidirectional Rotation:
 Suppression of back isomerization (dashed arrow in Figure~\ref{fig:scheme1}), ensuring unidirectional rotation. This can be achieved by selective excitation of the stable conformers and therefore requires the stable and metastable conformers to exhibit well separated absorption bands. (\crit{2})
    \item Preserved Photoreactivity: 
 High similarity of the transition density of both stable conformers to the $\pi-\pi^\ast$ character of the unsubstituted MNM, ensuring high EZ-isomerization probability, as indicated by the {\it photoreactivity score} (PRS), defined in Methods and Computational Details. (\crit{3})
\end{enumerate}

According to the criteria, we identified the most suitable isomer pairs.

\

\begin{figure}[H]
    \centering
    \includegraphics[width=7in]{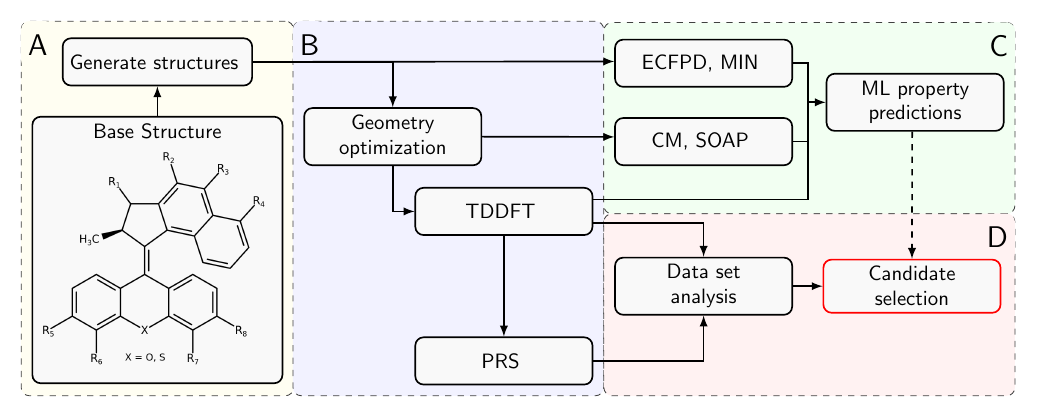}
    \caption{Flowchart of the full process of this study. First (A), the data set structures are generated from a base structure. Next (B), the generated structures are optimized with DFT, TDDFT calculations are performed, and PRS are calculated. In (C), the structures are encoded using molecular descriptors and passed to the machine learning models for predictions. Lastly (D), the data is analyzed and candidates are chosen according to our selection criteria. A dashed line indicates the potential use of ML predictions for candidate selection.}
    \label{flowchart}
\end{figure}

Lastly, to investigate the possibility to scale up our procedure to a larger chemical space, we explore the possibility to apply machine learning (ML), intended to replace costly quantum chemical predictions of the target quantities. To this end, three distinct ML models were trained on the calculated excited state molecular properties and a benchmark study of the models in combination with four different molecular representations was performed. 
We use physical molecular descriptors (CM and SOAP), requiring the three-dimensional structure of the molecules, as well as connectivity-based descriptors (ECFPD and MIN) (Figure \ref{flowchart}). 
Connectivity-based descriptors have the potential to accelerate the screening of candidates, since they avoid the computationally intense quantum mechanical geometry optimizations.

\section{Methods and Computational Details}

\subsection{Dataset Generation}
To generate the data set for this study, we followed a systematic approach (Figure~\ref{flowchart}) to explore the structural and electronic properties of molecules with varying substituents. The base molecule (Figure~\ref{ds2g_mol}) was adopted from previous research\cite{augulis2009light}. Its structure exists in two distinct helical conformers:\cite{moss1996basic} the {\it (P)}-conformer and the {\it (M)}-conformer. If the absolute configuration of the chiral center (asterisk in Figures~\ref{fig:scheme1} and \ref{ds2g_mol}) is chosen as indicated in Figure~\ref{ds2g_mol}, the {\it (M)}-conformer corresponds to the stable conformer, whereas the {\it (P)}-conformer corresponds to the meta-stable conformer. Additionally, the molecule was modified to feature either oxygen or sulfur at position X, resulting in four distinct base structures: {\it (M)}-{\bf Base-O} (oxygen, {\it (M)}-conformer), {\it (P)}-{\bf Base-O} (oxygen, {\it (P)}-conformer), {\it (M)}-{\bf Base-S} (sulfur, {\it (M)}-conformer) , and {\it (P)}-{\bf Base-S} (sulfur, {\it (P)}-conformer).

Before we generated the substituted structures, we first optimized the unsubstituted structures ({\it (M)}-{\bf Base-O}, {\it (P)}-{\bf Base-O}, {\it (M)}-{\bf Base-S}, and {\it (P)}-{\bf Base-S}) using density functional theory (DFT), preserving the predefined helicity.
{\it (P)}- and {\it (M)}-conformers can further be classified into anti- and syn-conformers, depending on the conformation of the stator. However, anti-{\it (M)} and syn-{\it (P)} conformations were determined to be significantly more stable than the syn-{\it (M)} and anti-{\it (P)} conformations.\cite{Durbeej2015PCCP,lucia2025first} For the sake of simplicity, we therefore consider only anti-{\it (M)} and syn-{\it (P)} conformers in this study. 

Then, to generate the substituted molecules (Figure~\ref{flowchart}A),
we added one donor and one acceptor group to the base structures, yielding a preliminary structural model of the molecules. Eight positions in the base structure (R$_1$-R$_8$ in Figure~\ref{ds2g_mol}) were selected for the attachment of substituents. All possible combinations of one donor group and one acceptor group were generated systematically, while the remaining positions were filled with hydrogen atoms. A total of 3 donor and 3 acceptor groups were considered, resulting in 9 unique donor-acceptor pairs with a total of $504$ possible permutations across the 8 positions. Given the four base structures, the total number of molecules in the data set is therefore 2016.
The preliminary structural models were further optimized using DFT.

\subsection{Quantum Mechanical Calculations} 
All quantum chemistry calculations were performed with
TURBOMOLE\cite{turbomole78,balasubramani2020turbomole,franzke2023turbomole}, employing the 
def2-SV(P)\cite{weigend2005balanced} basis set. 
Geometry optimizations 
were performed 
using DFT with the resolution of identity \cite{Eichkorn1997} approximation and the PBE\cite{PBE1996PhysRevLett.77.3865} functional. 
Excited-state single-point calculations were performed with TDDFT\cite{Furche2002,parker2017quadratic} with the hybrid PBE0\cite{Perdew1996a,Adamo1999J.Chem.Phys.} functional. 
Excitation energies, oscillator strengths, and TPASs of the lowest two excited singlet states, S$_1$ and S$_2$, respectively, were computed. We only considered symmetric TPA, where both incident photons have the same energy.

\subsection{Determination of the Photoreactivity Score}
A previous study\cite{lucia2025first} has shown, that the unsubstituted structures undergo EZ-isomerization upon excitation into
 S$_1$, corresponding to a $\pi-\pi^\ast$ transition. However, this may not hold for the substituted structures, as the substituents may alter the character of the excited states.
 To prevent the molecule from being trapped in an unreactive state, it is required that S$_1$ adopts a $\pi$ -- $\pi^*$ character, allowing the system to relax energetically toward a S$_1$-S$_0$ conical intersection, accompanied by a rotation of the central dihedral angle (C9a-C9-C1'-C1'a in Figure~\ref{ds2g_mol}). 

\begin{figure}[H]
    \centering
    \includegraphics[width=3.3in]{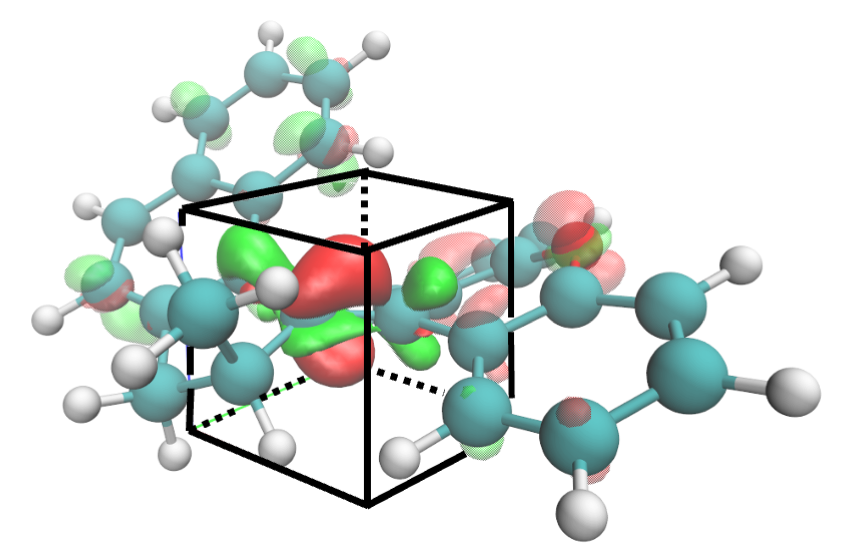}
    \caption{Transition density of S$_1$ $(\pi-\pi^*)$ of the {\it (M)-}{\bf Base-O} (density isovalue of 0.002). The density within the central box (solid iso-surface) was used to determine the PRS. The density outside of the box (transparent iso-surface) was discarded in the computation of the PRS. The graphic was created with VMD \cite{HUMP96}. A schematic of a box was overlayed for illustrative purposes (black).}
    \label{box_cut_example}
\end{figure}

To determine whether an electronic transition is likely to be a $\pi$ - $\pi^\ast$ transition leading to an EZ-photoisomerization, we compare the transition density of S$_1$ to the 
S$_1$ ($\pi-\pi^*$) transition density of the unsubstituted base structures.
To focus our analysis on the significant changes in the transition density relative to the base molecule, we defined a box around the central double bond (C1'=C9, defined in Figure~\ref{ds2g_mol}) and only considered the part of the transition density within the region of this box (Figure~\ref{box_cut_example}). 
The projection of the two transition densities within this region was used to calculate a continuous similarity score, which we called the PRS, the formula for which is given in Equation~\ref{eq:prs}. 

\begin{equation}\label{eq:prs}
    \text{PRS}_i=\frac{1}{2}\left( 1+\frac{{\textbf b}_\text{base}\cdot \textbf{b}_{i}}{|\textbf{b}_\text{base}||\textbf{b}_{i}|}\right)
\end{equation}

Where PRS$_i$ is the photoreactivity score of molecule $i$, {\bf b}$_{i}$ is the vector collecting all grid points of the transition density within the box of molecule $i$, and {\bf b}$_\text{base}$ the analogous vector for the base molecule. The boxes contain 32 $\times$ 32 $\times$ 32 grid points.
The PRS ranges from 0, corresponding to a perfect disagreement, to 1, a perfect match of the densities of the molecule and the base molecule.  

\subsection{Machine Learning Details}
We studied the ability to predict the photochemical properties of the molecules in our data set using three different ML architectures and encoded the chemical information using three commonly used molecular descriptors.

The models used in this study are i) Kernel Ridge Regression\cite{Rupp2011FastAA, Ramakrishnan2015} (KRR), ii) eXtreme Gradient Boosting\cite{Chen2016XGBoostAS} (XGB), and iii) Neural Network (NN)\cite{dl_tax}. 

\begin{enumerate}
  \item KRR: A regularized regression method using the kernel-trick. Implemented using the Python scikit-learn package\cite{scikit-learn}.
  \item XGB: An open source library for a gradient-boosted decision tree ensemble method using a continuous score for predictions.
  \item NN: Three dense layers connected via non-linear functions. Implemented with pytorch\cite{pytorch2019} and hyperparameter tuning performed with Ray Tune\cite{liaw2018tune}.
\end{enumerate}

The molecular descriptors used are i) Coulomb matrix\cite{mol_rep2022} (CM), ii) Smooth Overlap of Atomic Positions\cite{SOAP2013} (SOAP), and iii) Extended-connectivity fingerprints\cite{rogers2010ecfp} combined with RDKit descriptors\cite{rdkitdocs} (ECFPD).

\begin{enumerate}
    \item CM: Simple electrostatic atom-paired matrix. Implemented with the Python DScribe package\cite{Dscribe2020}. Total number of unique features: 1,770.
    \item SOAP: Fully differentiable local environment encoding of atoms. Implemented with the Python DScribe package\cite{Dscribe2020}. Total number of unique features: 37,170.  
    \item ECFPD: A combination of a two-dimensional fingerprint algorithm, molecular descriptors included in the open-source library RDKit, and custom structure descriptors. Total number of unique features: 4,325.
\end{enumerate}

In addition to standard molecular descriptors, we developed a minimal binary encoding scheme (MIN) for our molecules. This encoding uses 24 unique features to describe each molecule: 1 bit indicating X=O/X=S, 1 bit indicating helical conformation {\it (M)} or {\it (P)}, 6 bits for the presence of each substituent, 8 bits for the occupied donor positions, and 8 bits for the occupied acceptor positions. This set of features provides the minimal, yet complete information required to distinguish all molecules in our dataset.

While the physical descriptors CM and SOAP require the three-dimensional structure of the molecules, ECFPD and MIN, in contrast, are only derived from connectivity-based encoding (Figure \ref{flowchart}). 
However, limited structural information, such as helical conformation, is also contained in these descriptors. Additionally, the inclusion of RDKit molecular descriptors as features in ECFPD provides 125 features corresponding to physicochemical and physical properties and 85 fragment-based features describing counts of specific atom types, aromatic rings, or functional groups.\cite{rdkitdocs}


\section{Results and Discussion}
\subsection{Data Set Analysis}
We generated the set of 2016 molecules as described above and performed geometry optimizations. Cartesian coordinates of all compounds are available online together with the photochemical properties and transition densities\cite{zenododata}. Geometry optimization failed for 32 molecules, due to sterical interactions of donor/acceptor groups, in cases where both functional groups are located on the rotor. We removed these molecules (Table S1, SI) from the dataset, reducing the total number of molecules to 1984.
Of the 1984 stable molecules, 400 have donor and acceptor located on the rotor; 432 have both substituents located on the stator; the remaining 1152 candidates have one substituent on the rotor and the other substituent on the stator. For the latter two sets of molecules, the isomer pairs appearing in one rotational cycle are chemically different.
In total, we have 596 unique isomer pairs, describing 596 unique rotational cycles. 

Using TDDFT, we calculated excitation energies \exe{}, oscillator strengths \osc{}, and TPAS \tpas{} for S$_1$ (\exe{1}, \osc{1} and \tpas{1}, respectively) and S$_2$ (\exe{2}, \osc{2} and \tpas{2}, respectively) of all 1984 compounds. Three distinct variations of TPAS were computed: linear parallel TPAS, linear orthogonal TPAS, and circularly polarized TPAS. For simplicity, the subsequent analysis is exclusively focused on the linear parallel TPAS due to the significant correlations with linear orthogonal and circularly polarized TPAS (Figure S1, SI).

In addition to the 1984 substituted molecules, we also computed photochemical properties of nine reference compounds (Table~\ref{ref_mols}). These include the four unsubstituted base structures {\it (M)-}{\bf Base-O}, {\it (P)-}{\bf Base-O}, {\it (M)-}{\bf Base-S}, and {\it (P)-}{\bf Base-S}, 
and five compounds from the literature. The latter are comprised by four second-generation oxindole-based molecular motors, {\bf E$_S$-1}, {\bf E$_S$-2}, {\bf E$_S$-3}, and {\bf E$_S$-4} (Guinart et al.\cite{GUINART2024115649}), which exhibit two-photon absorption in the NIR, and {\bf Motor-3} (García-López et al.\cite{garcia2017molecular}), which has been shown to disrupt cellular bilayers via NIR TPA.\cite{liu2019near}
The most stable conformer of each compound was chosen, and the names of the reference compounds were left unchanged from the original publications.

\begin{table}[H]
    \centering
    \begin{tabular}{l|ccr|ccr}
        Molecule & \exe{1} [eV] & \osc{1} [a.u.] & \tpas{1} [a.u.] & \exe{2} [eV] & \osc{2} [a.u.] & \tpas{2} [a.u.] \\
    \hline
        {\it (M)-}{\bf Base-O} & 2.94 & 0.386 &  179  & 3.71 & 0.010 & 10,276 \\
        {\it (P)-}{\bf Base-O} & 2.46 & 0.358 &  328  & 3.45 & 0.019 & 15,545 \\
        {\it (M)-}{\bf Base-S} & 3.34 & 0.299 &  163  & 3.76 & 0.063 & 2,452 \\
        {\it (P)-}{\bf Base-S} & 2.49 & 0.339 & 1,307 & 3.37 & 0.018 & 6,949 \\
    \hline
        {\bf E$_S$-1}          & 2.90 & 0.275 &   57  & 3.25 & 0.377 & 3,408 \\
        {\bf E$_S$-2}          & 2.93 & 0.384 &  112  & 3.26 & 0.007 &  19 \\
        {\bf E$_S$-3}          & 2.84 & 0.413 &  206  & 3.18 & 0.269 & 4,647 \\
        {\bf E$_S$-4}          & 2.87 & 0.521 &  465  & 3.26 & 0.142 & 6,443 \\
    \hline
        {\bf Motor-3}          & 3.28 & 0.228 &  308  & 3.63 & 0.113 & 1,946 \\
    \end{tabular}
    \caption{Photochemical properties of reference compounds.}
    \label{ref_mols}
\end{table}

\subsubsection{Data Distribution}
\begin{figure}[H]
    \centering
    \includegraphics{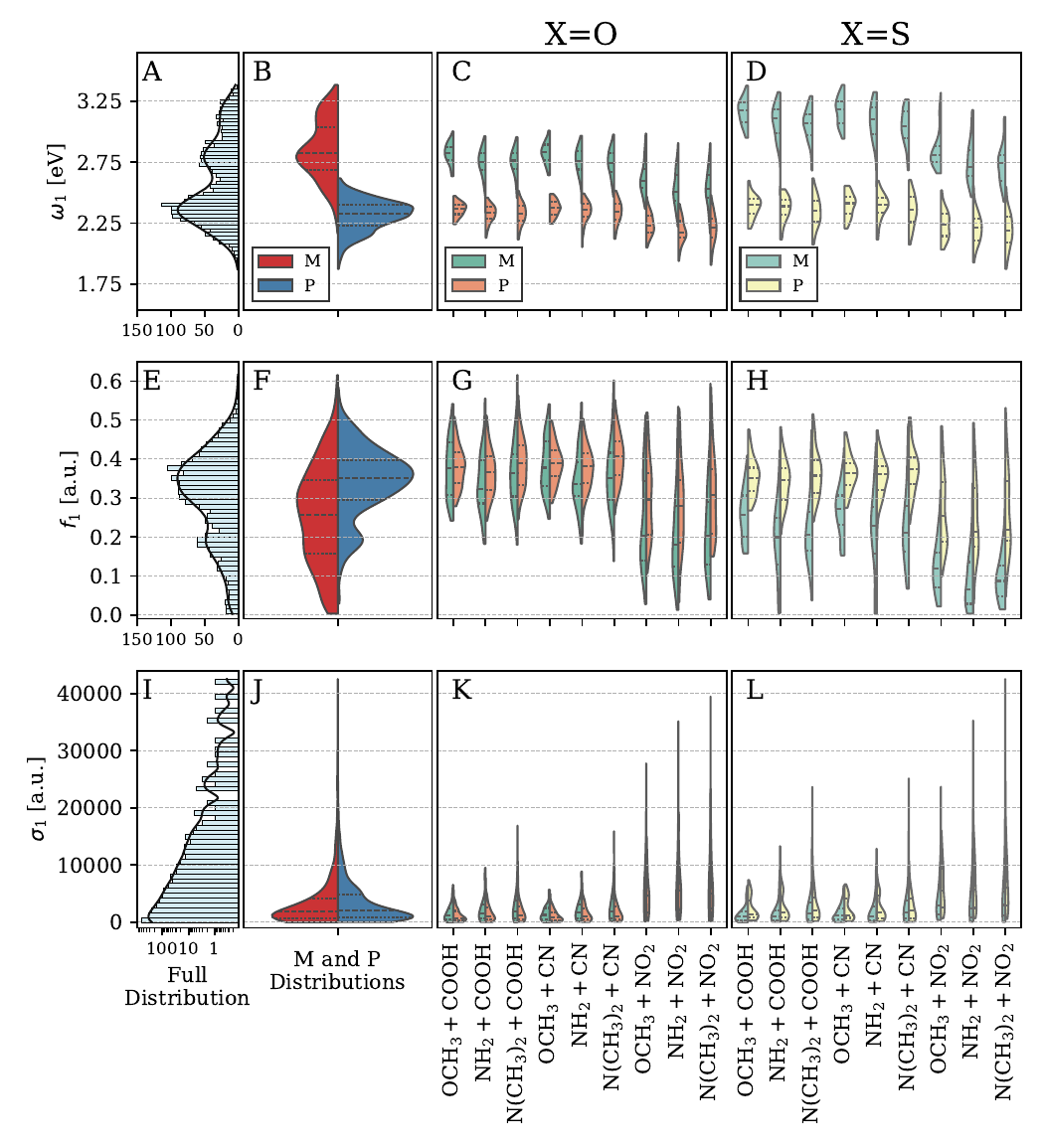}
    \caption{Histograms of \exe{1} (A), \osc{1} (E), \tpas{1} (I, logarithmic scale), combined with violin plots of the distributions of {\it (M)}- and {\it (P)}-conformers (B, F, J respectively), donor-acceptor pairings for X=O (C, G, K respectively) and for X=S (D, H, L respectively).}
\label{hist_and_da_pair_violin_split_E_TPAS_Oszi_1st}
\end{figure}

The S$_1$ excitation energies of the 1984 generated compounds range from 1.87 to 3.38 eV (Figure~\ref{hist_and_da_pair_violin_split_E_TPAS_Oszi_1st}A), with a median \exe{1}  of 2.49 eV. {\it (P)}-conformers exhibit smaller \exe{1} than {\it (M)}-conformers (Figure~\ref{hist_and_da_pair_violin_split_E_TPAS_Oszi_1st}B); this is true for all molecules in our data set (Figure S2A, SI).  This is also reflected in the violin plots for the distinct donor-acceptor pairs (Figure~\ref{hist_and_da_pair_violin_split_E_TPAS_Oszi_1st}C,D), both for X=O and X=S. However, the separation of the \exe{1}-distributions of {\it (M)}- and {\it (P)}-conformers is larger for X=S. This is mainly due to sulfur-containing {\it (M)}-conformers exhibiting larger \exe{1} than their oxygen-containing counterparts. The distribution of molecules containing \ce{NO2} as acceptor shows smaller excitation energies and larger overlap between the distributions of {\it (M)}- and {\it (P)}-conformers than molecules with COOH- or CN-acceptor groups. 

The S$_1$ oscillator strengths range from 0.00298 to 0.616 a.u., with a median \osc{1} of 0.315 a.u.~(Figure~\ref{hist_and_da_pair_violin_split_E_TPAS_Oszi_1st}E). 
 \osc{1}-distributions of {\it (M)}- and {\it (P)}-conformers are less separated than \exe{1}-distributions. On average, oscillator strengths are smaller for {\it (M)}-conformers than for {\it (P)}-conformers (Figure~\ref{hist_and_da_pair_violin_split_E_TPAS_Oszi_1st}F). The mean \osc{1} is larger for X=O than for X=S (Figure~\ref{hist_and_da_pair_violin_split_E_TPAS_Oszi_1st}G,H). As in the case of \exe{1}, 
 molecules with \ce{COOH}- and \ce{CN}-acceptors show similar \osc{1}-distributions, while \ce{NO2}-substituted molecules exhibit a distribution shifted towards smaller values. Interestingly, we also notice that the maximum \osc{1}-value increases with decreasing Hammett parameter (given in Figure~\ref{ds2g_mol}) of the donor groups.

The S$_1$ TPAS-values range from 2.84 to 44,000 a.u.~(Figure~\ref{hist_and_da_pair_violin_split_E_TPAS_Oszi_1st}I), with a median of 1903 a.u. The number of molecules decays exponentially with increasing TPAS, with only a few tens of molecules exhibiting values above 20,000 a.u. Distributions of {\it (M)}- and {\it (P)}-conformers show no significant difference (Figure~\ref{hist_and_da_pair_violin_split_E_TPAS_Oszi_1st}J). The largest \tpas{1}-values are found for molecules with \ce{NO2} (Figure~\ref{hist_and_da_pair_violin_split_E_TPAS_Oszi_1st}K,L),
with maximum values above 30,000 a.u. As in the case of \osc{1}, a decreasing Hammett parameter leads to an increased maximum value for \tpas{1}.

\begin{figure}[H]
    \centering
    \includegraphics{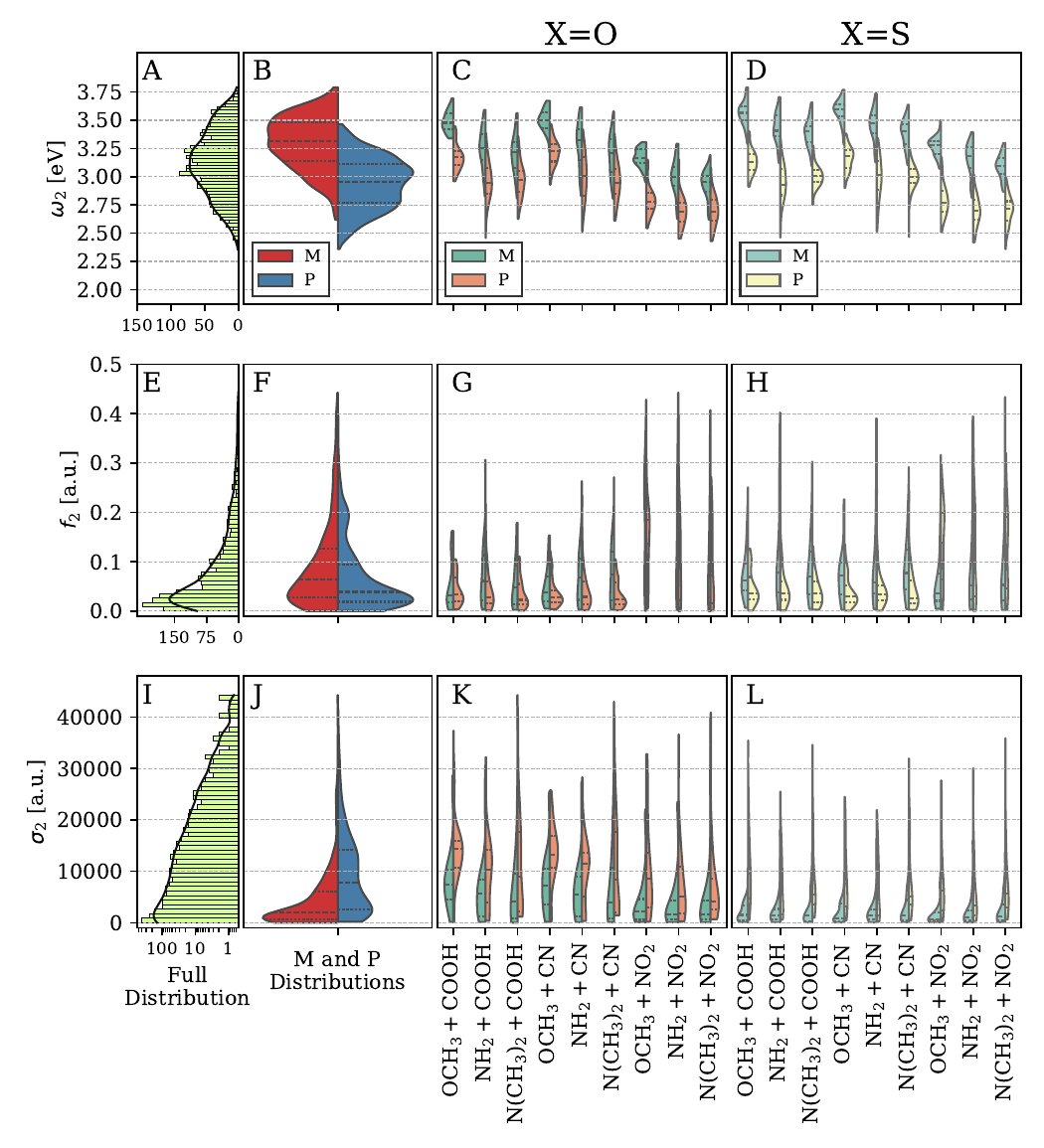}
    \caption{Histograms of \exe{2} (A), \osc{2} (E), \tpas{2} (I, logarithmic scale), combined with violin plots of the distributions of {\it (M)}- and {\it (P)}-conformers (B, F, J respectively), donor-acceptor pairings for X=O (C, G, K respectively) and for X=S (D, H, L respectively).}
    \label{hist_and_da_pair_violin_split_E_TPAS_Oszi_2nd}
\end{figure}

The S$_2$ excitation energies range from 2.36 eV to 3.79 eV, with a Gaussian-like distribution centered around 3.12 eV (Figure~\ref{hist_and_da_pair_violin_split_E_TPAS_Oszi_2nd}A). Molecules in {\it (M)}-conformation have larger \exe{2} on average than {\it (P)}-conformers (Figure~\ref{hist_and_da_pair_violin_split_E_TPAS_Oszi_2nd}B). Distributions of \exe{2} for a given donor-acceptor pair
(Figure~\ref{hist_and_da_pair_violin_split_E_TPAS_Oszi_2nd}C,D) are similar for X=O and X=S. 

In contrast to \osc{1}, the \osc{2}-distribution decays exponentially with increasing values (Figure~\ref{hist_and_da_pair_violin_split_E_TPAS_Oszi_2nd}E). The majority (72\%) of the \osc{2}-values are below 0.1 a.u., while only 7\% of the \osc{1}-values are below 0.1 a.u. 
\osc{2}-distributions for {\it (M)}- and {\it (P)}-conformers are more similar to each other than in the case of \osc{1}.
Molecules with X=O have on average smaller \osc{2} than molecules with X=S (Figure~\ref{hist_and_da_pair_violin_split_E_TPAS_Oszi_2nd}G,H). 
No clear correlations between \osc{2} and the Hammett parameters are visible.

Similar to S$_1$, the number of molecules decays exponentially
with increasing \tpas{2}, but more molecules are found with values above 20,000 a.u.
(Figure~\ref{hist_and_da_pair_violin_split_E_TPAS_Oszi_2nd}I).  The number of {\it (M)}-conformers decays faster with increasing \tpas{2}, than the number of {\it (P)}-conformers (Figure~\ref{hist_and_da_pair_violin_split_E_TPAS_Oszi_2nd}J). For each donor-acceptor pair, the maximum \tpas{2}-values are larger for X=O than for the corresponding pair with X=S (Figure~\ref{hist_and_da_pair_violin_split_E_TPAS_Oszi_2nd}K,L). Acceptor and donor strength do not have an obvious effect on \tpas{2}.

\begin{figure}[H]
    \centering
    \includegraphics{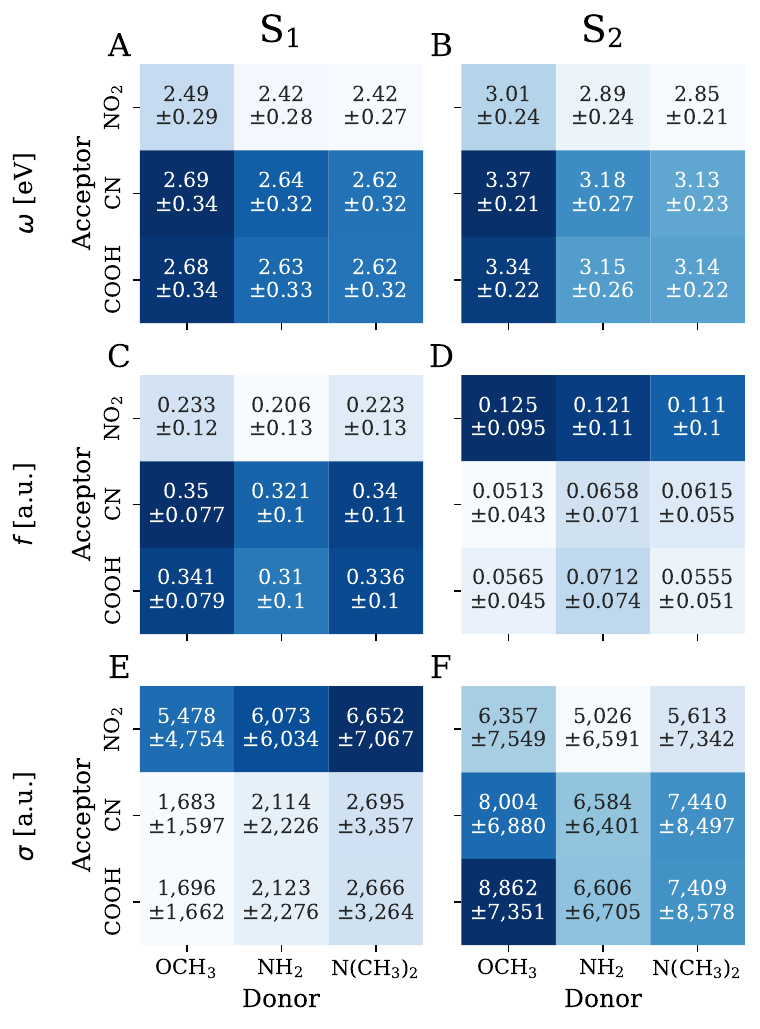}
    \caption{Mean value and standard deviation of S$_1$ and S$_2$ properties, \exe{}, \osc{}, and \tpas{}, as a function of donor/acceptor pair.}
    \label{heatmap_da_groups_ExcE_TPAS_Oszi_mean_std}
\end{figure}

Inspecting the mean values for each property as a function of the specific acceptor (Figure~\ref{heatmap_da_groups_ExcE_TPAS_Oszi_mean_std}), we see that COOH and CN lead to similar \exe{1} (panel A)
and \exe{2} (panel B) values for a given donor, whereas \ce{NO2} results in  consistently smaller values.
For \osc{1} (panel C), \ce{NO2} also leads to significantly smaller values than COOH and CN, but the opposite is true for \osc{2} (panel D).
For \tpas{}, in contrast, \ce{NO2} leads to significantly larger values in S$_1$ (panel E) and significantly smaller values in S$_2$ (panel F). 
Across all properties, we see that the average values of \ce{COOH} and \ce{CN} substituent molecules are almost identical, while the average values of \ce{NO2} substituent molecules differ noticeably.  

For both excitations, increasing donor strength leads to a decrease in energy (panels A and B).
No clear correlation is seen between the donor strength and the oscillator strength for both excitations (panels C and D, respectively).
For \tpas{1}, increasing donor strengths leads to increased values (panel E). For \tpas{2}, however, no clear correlation is visible (panel F).

\subsubsection{Donor-Acceptor Position Effects}
 To gain a more detailed understanding of our data set, we studied the dependence of the different photochemical properties as a function of the donor-acceptor distance.

\begin{figure}[H]
    \centering
    \includegraphics[width=3.3in]{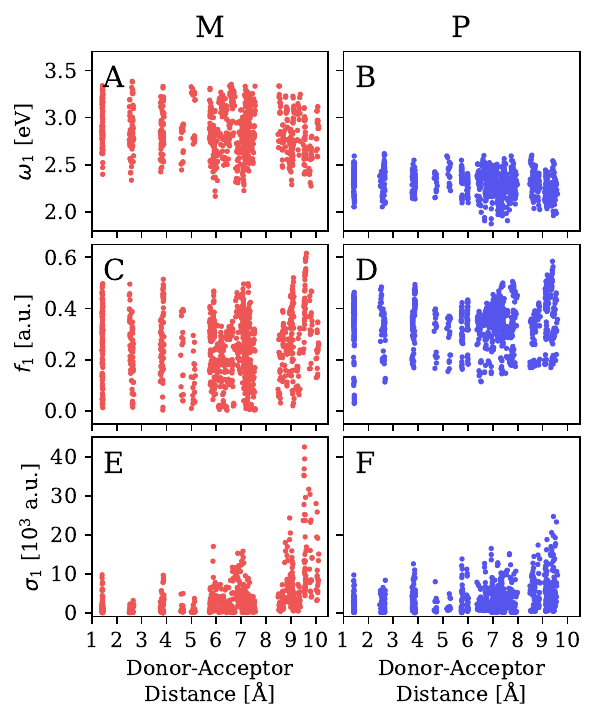}
    \caption{\exe{1}, \osc{1} and \tpas{1} as a function of donor-acceptor distance, given for {\it (M)}- and {\it (P)}-conformers.}
    \label{props_vs_dadist_mp}
\end{figure}

Apart from a minor decrease in its mean value at larger distances, \exe{1} is essentially independent of the donor–acceptor distance (Figure~\ref{props_vs_dadist_mp}A,B). For \osc{1}, no obvious trend is visible (Figure~\ref{props_vs_dadist_mp}C,D). 

In contrast, for \tpas{1}, we observe more molecules with large values for larger donor-acceptor distances (Figure~\ref{props_vs_dadist_mp}E,F). This effect is significantly more pronounced for the {\it (M)}-conformers than for the {\it (P)}-conformers. 
This is consistent with the positive relationship between TPAS and the size of the $\pi$-electron system between donor and acceptor, which is well-documented.\cite{TPADesign2009Pawlicki}

\begin{figure}[H]
    \centering
    \includegraphics{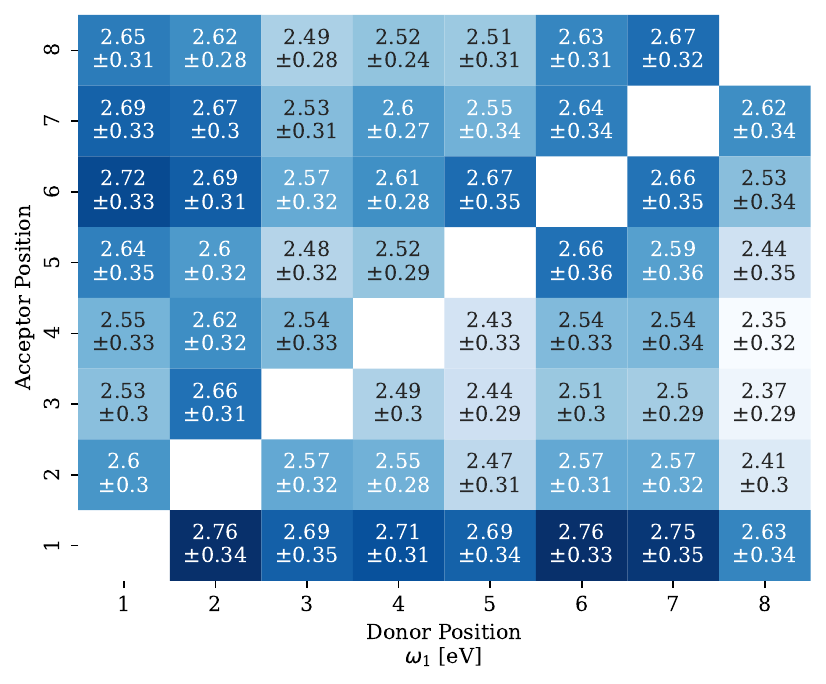}
    \caption{Mean value and standard deviation of \exe{1} as a function of donor/acceptor positions.}
    \label{heatmap_da_positions_1ExceE_mean_std}
\end{figure}

To gain further insights, we also investigated the dependence of the properties on the specific positions (R$_1$-R$_8$, Figure~\ref{ds2g_mol}) of donor and acceptor. 
For \exe{1}, mean values are not significantly affected by the donor/acceptor positions, except that we find generally higher values when the acceptor occupies R$_1$ (Figure~\ref{heatmap_da_positions_1ExceE_mean_std}).
This is possibly due to R$_1$ being the only position that is not connected to the $\pi$-electron system via conjugated double bonds.

\begin{figure}[H]
    \centering
    \includegraphics{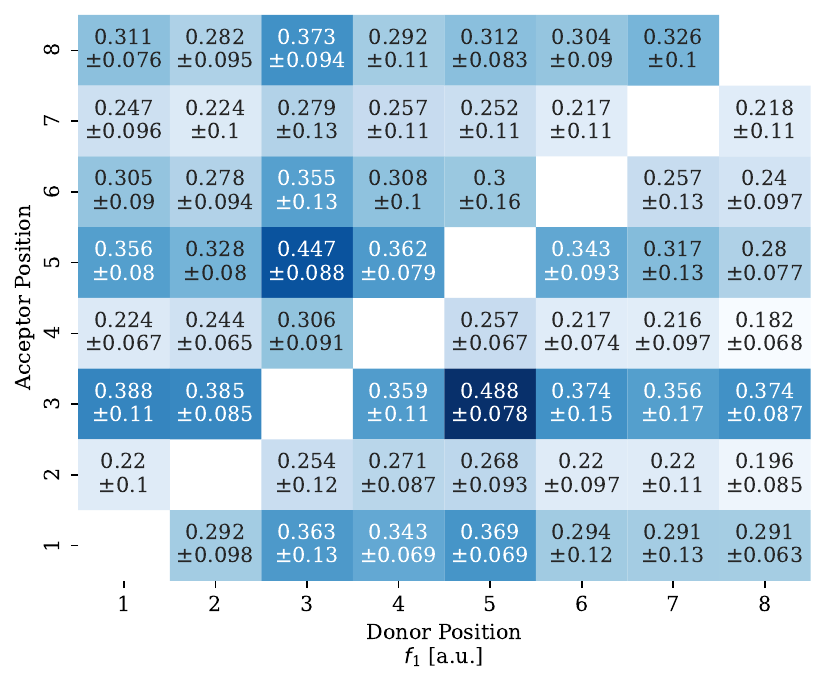}
    \caption{Mean value and standard deviation of \osc{1} as a function of donor/acceptor positions.}
    \label{heatmap_da_positions_1Osz_mean_std}
\end{figure}

For \osc{1}, in contrast, we see an interesting relationship between its mean value and the donor/acceptor positions (Figure~\ref{heatmap_da_positions_1Osz_mean_std}):  We find that, regardless of the donor, \osc{1} values are on average larger if the acceptor occupies positions R$_1$, R$_3$, R$_5$, R$_6$, and R$_8$. The highest mean values occur when the donor and acceptor occupy positions R$_3$ and R$_5$, respectively, or vice versa.

\begin{figure}[H]
    \centering
    \includegraphics{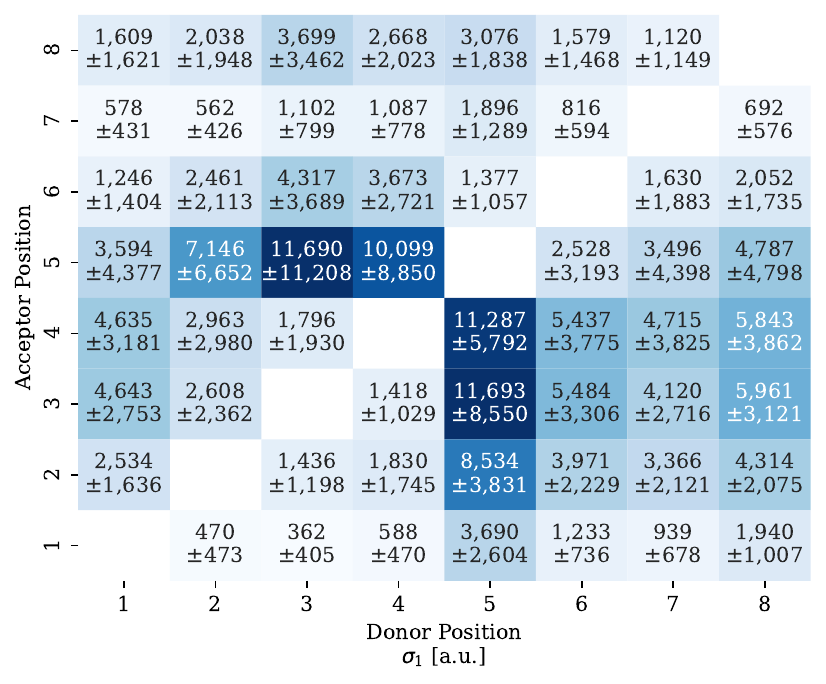}
    \caption{Mean value and standard deviation of  \tpas{1} as a function of donor/acceptor positions.}
    \label{heatmap_da_positions_1TPAS_mean_std}
\end{figure}

Compared to \osc{1}, \tpas{1} values are even more sensitive to the specific donor/acceptor positions (Figure~\ref{heatmap_da_positions_1TPAS_mean_std}). We observe the highest values (7,000 -- 12,000 a.u.) if either the donor is at R$_5$ and the acceptor occupies R$_2$, R$_3$, or R$_4$, or vice versa. The large mean values for acceptor position R$_5$ are mainly caused by molecules with \ce{NO2} (Figure S4--S6, SI).
In contrast, the presence of acceptor groups at positions R$_1$ and R$_7$ leads to the lowest \tpas{1} values, regardless of the selected donor position. 

\subsubsection{Analysis of the Photoreactivity Score}
For each molecule in the data set, we computed the PRS.
As we are only interested in the forward isomerization, we only examine the PRS of the stable {\it (M)}-conformers.
For the {\it (M)}-conformers, the calculated PRS range from 0.0659 to 0.996 with an average score of $0.72\pm 0.3$.

\begin{figure}[H]
    \centering
    \includegraphics{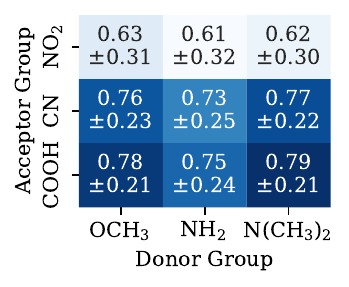}
    \caption{Mean value and standard deviation of the PRS as a function of the donor/acceptor pair S$_1$.}
    \label{dot_group_heatmap}
\end{figure}

Analyzing the effect of the chemical nature of the donor/acceptor pair on the PRS (Figure~\ref{dot_group_heatmap}), we notice that molecules with \ce{NO2} have significantly lower average PRS values than molecules with \ce{COOH} or \ce{CN} as acceptor. 
This indicates that on average, molecules with \ce{NO2} are less likely to exhibit $\pi$ -- $\pi^*$ character in S$_1$ than in the case of the other two acceptor groups.
No clear trend between the donor type and the PRS can be observed. 

\begin{figure}[H]
    \centering
    \includegraphics{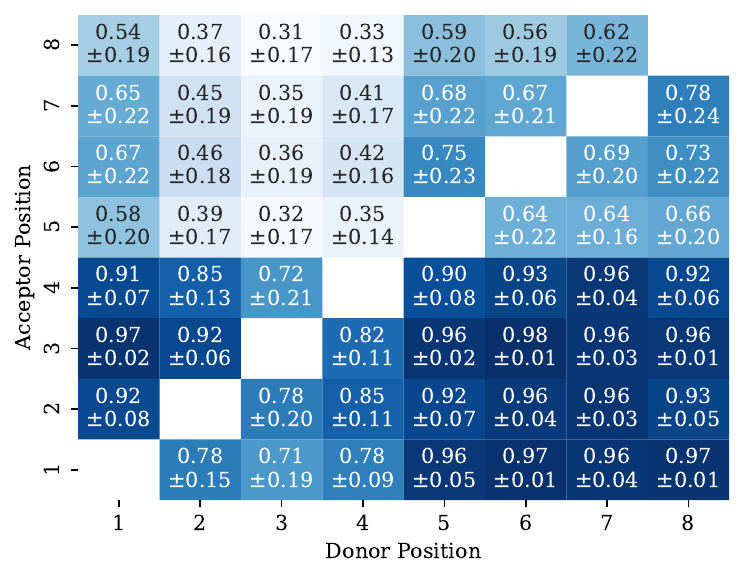}
    \caption{Mean value and standard deviation of the PRS as a function of donor/acceptor position for S$_1$.}
    \label{dot_position_heatmap}
\end{figure}

 Inspecting the mean value of the PRS as a function of donor/acceptor positions (Figure~\ref{dot_position_heatmap}), we observe values ranging from 0.31 to 0.98. 
 We observe the largest values ($>$0.90) for molecules with the acceptor at positions R$_1$--R$_4$ (located on the rotor) and donor at positions R$_5$--R$_8$ (located on the stator). The opposite arrangement (acceptor on stator and donor on rotor), in contrast, leads to the opposite effect: smallest values are found in the upper left quadrant of Figure~\ref{dot_position_heatmap}. For the case where both acceptor and donor are located on the rotor, we find relatively high PRS values (0.71--0.97), whereas significantly smaller values (0.56--0.78) are found when both substituents are located on the stator.

 In view of the low average values for \ce{NO2} (Figure~\ref{dot_group_heatmap}), we conclude 
 that the occupied positions have significantly more impact on the PRS than the chemical nature of the substituent.
 This is further supported by a detailed analysis of the PRS for each molecule (Figures S7-S15, SI).

\subsubsection{Selection of Isomer Pairs}
The isomer pairs in our data set describe a total of 596 rotational cycles. According to the criteria defined above, we now select isomer pairs that are most promising to achieve high quantum yields for potential nanomotor applications. 

\crit{2} (selective excitation of stable conformers) and \crit{3} (PRS close to 1) each excludes a portion of the data set unsuited for our purpose. \crit{1} (efficient absorber) ranks the remaining isomer pairs.

\begin{figure}[H]
    \centering
    \includegraphics{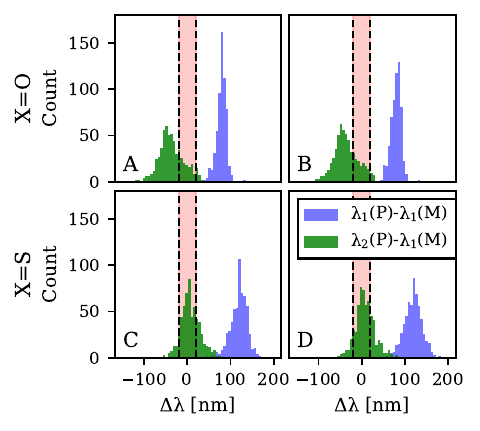}
    \caption{Histograms of the wavelength differences $\Delta \lambda$ for each isomer pair. $\lambda_1(M)$ denotes the wavelength of \exe{1} of the {\it (M)}-conformer,
    $\lambda_1(P)$ denotes the wavelength of  \exe{1} of the {\it (P)}-conformer, and
    $\lambda_2(P)$ denotes the wavelength of \exe{2} of the {\it (P)}-conformer. Panels A and C show $\Delta \lambda$ within the same isomer, wheres panels B and D show $\Delta \lambda$ between the two corresponding isomers within one isomer pair. Vertical lines (black) were added at positive and negative differences of 20 nm with the enclosed range highlighted in red.}
    \label{plt_exc_diff}
\end{figure}

Addressing \crit{2}, 
we inspect the differences in excitation wavelength $\Delta \lambda$ of {\it (M)}- and {\it (P)}-conformers of each isomer pair (Figure~\ref{plt_exc_diff}). If $\omega_1$ or $\omega_2$ of the {\it (P)}-conformers are too close to $\omega_1$ of their corresponding {\it (M)}-conformer, it will not be possible to excite the {\it (M)}-conformer selectively. 
Based on the typical linewidth of experimental absorption bands of overcrowded alkene nanomotors\cite{vicario2005controlling}, we estimate that a minimum threshold difference $\Delta \lambda_\text{min}$ of $20\;\mathrm{nm}$ is necessary to allow for selective excitation of the conformers (red interval in Figure~\ref{plt_exc_diff}). For a given isomer pair ({\bf A}/{\bf B}), one must evaluate $\Delta \lambda$ between {\it (M)}-{\bf A} and {\it (P)}-{\bf A}, and $\Delta \lambda$ between {\it (M)}-{\bf B} and {\it (P)}-{\bf B} (Figure~\ref{plt_exc_diff}A,C).
In addition, one also must evaluate $\Delta \lambda$ between {\it (M)}-{\bf A} and {\it (P)}-{\bf B} and $\Delta \lambda$ between {\it (M)}-{\bf B} and {\it (P)}-{\bf A} (Figure~\ref{plt_exc_diff}B,D). 

For all isomer pairs, we see that $\lambda_1(P)-\lambda_1(M)$ is always above 20 nm (blue in Figure~\ref{plt_exc_diff}).
For $\lambda_2(P)-\lambda_1(M)$, however, 28\% of the isomer pairs with X=O and 76\% of the pairs with X=S are below 20 nm (green in Figure~\ref{plt_exc_diff}).
For these molecules, selective excitation cannot be guaranteed.
In summary, \crit{2} leads to the elimination of 52\% of the isomer pairs.

\begin{figure}[H]
    \centering
    \includegraphics{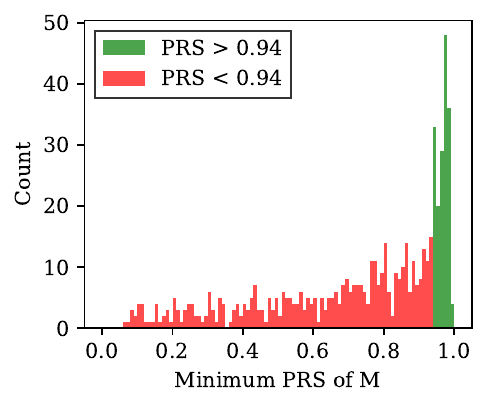}
    \caption{Histogram of the lower PRS of the two {\it (M)}-conformers for each isomer pair. 
    }
    \label{fig:PRS_distribution}
\end{figure}

Next, we filter our data set based on the PRS (\crit{3}). 
Small PRS values suggest that S$_1$ is unlikely to have $\pi-\pi^*$ character. 
The distribution of the PRS shows a marked increase in the number of molecules above 0.94 (Figure~\ref{fig:PRS_distribution}).
We visually inspected the transition densities of a considerable number of molecules and found that above 0.94 all investigated molecules showed a clearly defined $\pi-\pi^*$ transition. 
Therefore, 0.94 was chosen as a threshold to assign a $\pi-\pi^*$ character. 
This threshold must be met by both {\it (M)}-conformers of the isomer pairs. 
We find that the PRS of {\it (M)}-conformers of two corresponding isomers are highly correlated with a correlation coefficient of 0.98. 
\crit{3} excludes 72\% of the isomer pairs.

\crit{2} and \crit{3} combined, reduce the number of isomer pairs from 596 to 116, corresponding to an elimination of 81\%.

Finally, applying \crit{1}, we rank the remaining 19\% of the pairs according to their oscillator strength or TPAS, depending on whether they should be employed in one- or two-photon applications, respectively. To ensure that quantum yields of both forward photoisomerization steps (step 1 and step 3 in Figure~\ref{fig:scheme1}) are sufficiently high, we must consider both {\it (M)}-conformers of each isomer pair. 
To enable a ranking of the isomer pairs, we computed the geometric mean of the properties between the two {\it (M)}-conformers ($\tilde{f_{1}}$ and $\tilde{\sigma_{1}}$).

\begin{figure}[H]
\centering
\includegraphics[width=5in]{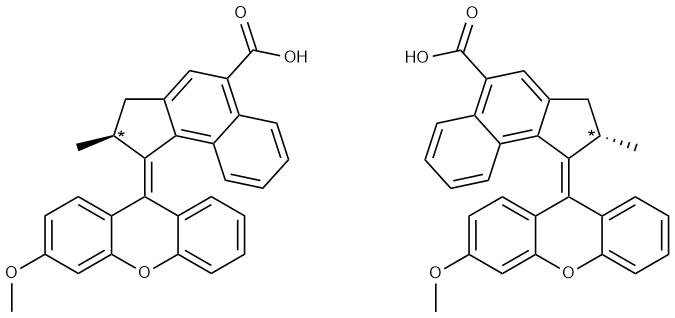}
\caption{Molecular structure of the best isomer pair {\bf (5-\ce{OCH3},3-\ce{COOH}/8-\ce{OCH3},3-\ce{COOH})-O} (Identification number\cite{zenododata}: 0251om, 0197om), ranked by $\tilde{f_{1}}$ and filtered against all selection criteria. Photochemical properties are summarized in Table~\ref{tab_best_osc}.}
\label{fig_best_osc}
\end{figure}

The 10 best isomer pairs ranked according to $\tilde{f_{1}}$ exhibit $\tilde{f_{1}}$-values ranging between 0.478 and 0.500 a.u.  
All 10 isomer pairs have X=O, which is expected as their maximum \osc{1} are generally larger than those with X=S (Figure~\ref{hist_and_da_pair_violin_split_E_TPAS_Oszi_1st}G,H). Half of them have CN as acceptor and the other half have COOH as acceptor, consistent with Figure~\ref{heatmap_da_groups_ExcE_TPAS_Oszi_mean_std}C. As donor, seven have \ce{OCH3}, two have \ce{NH2}, and one has \ce{N(CH3)2}. For all 10 pairs, the acceptor is located on the rotor at R$_3$, as expected from Figure~\ref{heatmap_da_positions_1Osz_mean_std}. 
Four isomer pairs also have the donor on the rotor, while the remaining pairs have the donor on the stator.
The best isomer pair in terms of $\tilde{f_{1}}$ is shown in Figure~\ref{fig_best_osc}; its properties are given in Table~\ref{tab_best_osc}. The remaining 9 best isomer pairs are given in the SI (Figure S18-26, SI).
Compared to most of the reference structures listed in Table \ref{ref_mols}, the best candidate shows an improved oscillator strength. Only {\bf E$_S$-4}, exhibits similar \osc{1}.

\begin{table}[H]
\centering
\begin{tabular}{c|cc|cc}
& \multicolumn{2}{c|}{\bf (5-\ce{OCH3},3-\ce{COOH})-O} & \multicolumn{2}{|c}{\bf (8-\ce{OCH3},3-\ce{COOH})-O} \\
Property    & {\it (M)} & {\it (P)} & {\it (M)} & {\it (P)} \\
\hline
$\mathit{\omega}_1$ [eV] & 2.66 & 2.26 & 2.63 & 2.25 \\
$\lambda_1$ [nm] & 466 & 548 & 471 & 552 \\
$\mathit{f}_1$ [a.u.] & 0.541 & 0.508 & 0.461 & 0.456 \\
$\mathit{\sigma}_1$ [a.u.] & 6,519 & 3,459 & 3,965 & 2,230 \\
\hline
$\mathit{\omega}_2$ [eV] & 3.43 & 3.23 & 3.38 & 3.20 \\
$\lambda_2$ [eV] & 362 & 384 & 366 & 387 \\
$\mathit{f}_2$ [a.u.] & 0.014 & 0.010 & 0.050 & 0.022 \\
$\mathit{\sigma}_2$ [a.u.] & 15,632 & 17,767 & 10,225 & 16,356 \\
\hline
PRS & 0.985 & 0.979 & 0.981 & 0.982 \\
\hline
\end{tabular}
\caption{Photochemical properties of the isomer pair shown in Figure~\ref{fig_best_osc}.}
\label{tab_best_osc}
\end{table}

  \par\medskip

\begin{figure}[H]
\centering
\includegraphics[width=5in]{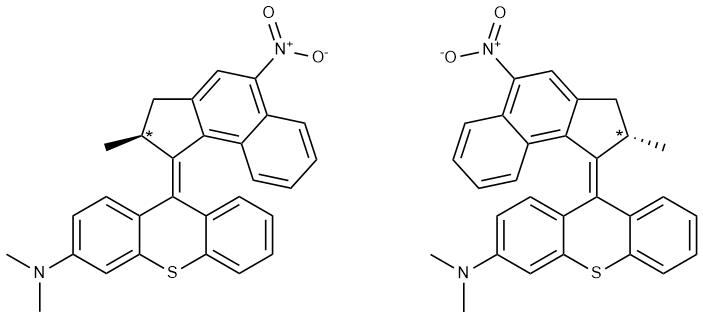}
\caption{Molecular structure of the best isomer pair {\bf (5-\ce{N(CH3)2},3-\ce{NO2}/8-\ce{N(CH3)2},3-\ce{NO2})-S} (Identification number\cite{zenododata}: 0243sm, 0189sm), ranked by $\tilde{\sigma_{1}}$ and filtered against all selection criteria. Photochemical properties are summarized in Table~\ref{tab_best_tpas}.}
\label{fig_best_tpas}
\end{figure}

Ranking the candidates according to $\tilde{\sigma_{1}}$, the 10 best isomer pairs exhibit $\tilde{\sigma_{1}}$-values ranging from 8,979  to 25,452 a.u. The best three have X=S, while the remaining seven have X=O. As acceptor, seven have \ce{NO2}, two CN, and one COOH. The preference of \ce{NO2} as acceptor is expected from the large average \tpas{1}-values (Figure~\ref{heatmap_da_groups_ExcE_TPAS_Oszi_mean_std}E). As donor, five have \ce{N(CH3)2}, four \ce{OCH3}, and one \ce{NH2}. All but one pair (R$_4$) have the acceptor at R$_3$; this is expected from the average \tpas{1}-values (Figure~\ref{heatmap_da_positions_1TPAS_mean_std}). Seven isomer pairs have the donor in position R$_5$/R$_8$, one in position R$_1$, one in position R$_2$ and lastly one in position R$_6$/R$_7$. 
All 10 pairs have S$_1$ two-photon absorption energies in the NIR. 
The best isomer pair in terms of $\tilde{\sigma_{1}}$ is shown in Figure~\ref{fig_best_tpas}; its properties are given in Table~\ref{tab_best_tpas}. The remaining nine best isomer pairs are given in the SI (Figure S27--35, SI).
Compared to its unsubstituted structure ({\it (M)-}{\bf Base-S}, Table~\ref{ref_mols}), the best candidate shows an increased \tpas{1} by about two orders of magnitude. Compared to the structures that have been used with TPA\cite{Liu2019ACSNano,GUINART2024115649} we achieve an increase by at least a factor of 30.

\begin{table}[H]
\centering
\begin{tabular}{c|cc|cc}
& \multicolumn{2}{c|}{\bf (5-\ce{N(CH3)2},3-\ce{NO2})-S} & \multicolumn{2}{|c}{\bf (8-\ce{N(CH3)2},3-\ce{NO2})-S} \\
Property    & {\it (M)} & {\it (P)} & {\it (M)} & {\it (P)} \\
\hline
$\mathit{\omega}_1$ [eV] & 2.59 & 2.06 & 2.45 & 1.98 \\
$\lambda_1$ [nm] & 480 & 603 & 505 & 628 \\
$\mathit{f}_1$ [a.u.] & 0.352 & 0.531 & 0.177 & 0.427 \\
$\mathit{\sigma}_1$ [a.u.] & 42,513 & 12,580 & 15,238 & 8,050 \\\hline
$\mathit{\omega}_2$ [eV] & 3.04 & 2.90 & 3.03 & 2.83 \\
$\lambda_2$ [eV] & 408 & 427 & 410 & 437 \\
$\mathit{f}_2$ [a.u.] & 0.013 & 0.067 & 0.105 & 0.072 \\
$\mathit{\sigma}_2$ [a.u.] & 6,508 & 13,133 & 2,998 & 7,936 \\\hline
PRS & 0.956 & 0.976 & 0.950 & 0.984 \\\hline
\end{tabular}
\caption{Photochemical properties of the isomer pair shown in Figure~\ref{fig_best_tpas}.}
\label{tab_best_tpas}
\end{table}

\subsection{Machine Learning Results}
\begin{figure}[H]
    \centering
    \includegraphics{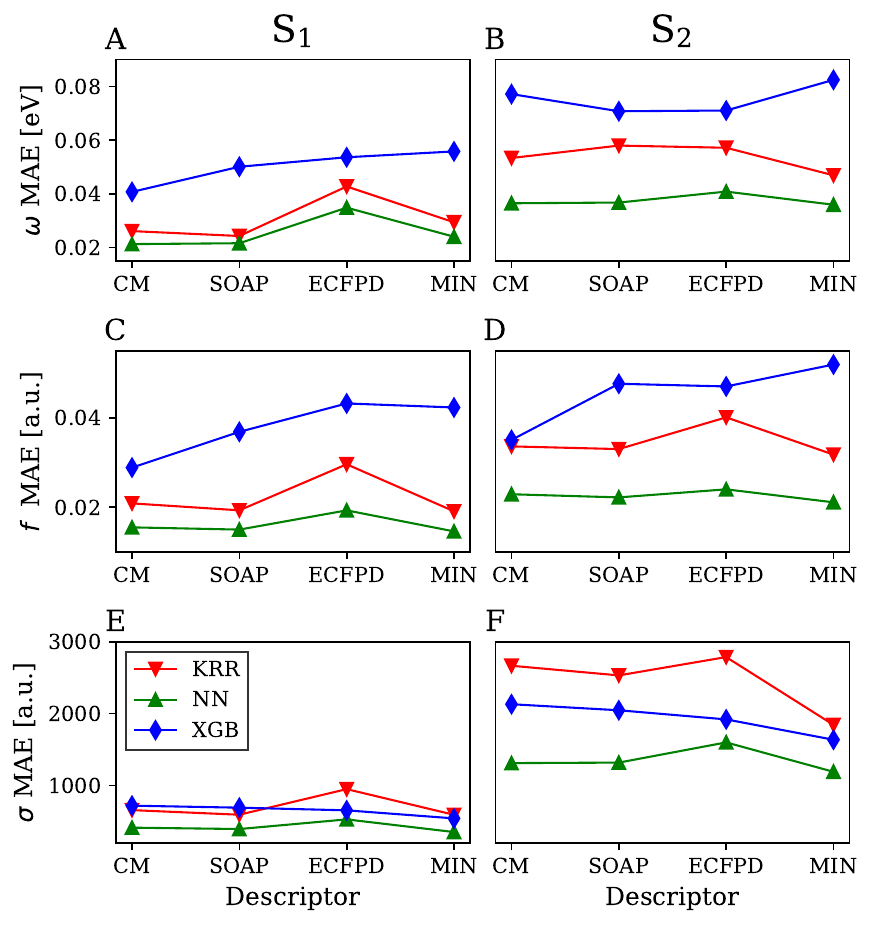}
    \caption{MAE of \exe{}, \osc{}, and \tpas{} for S$_1$ and S$_2$, as a function of ML model and molecular descriptor.}
    \label{MAE_plots}
\end{figure}

The mean absolute error (MAE) was used to evaluate the accuracy of the predictions for \exe{}, \osc{}, and \tpas{} of S$_1$ and S$_2$ with the three machine learning models and four molecular descriptors chosen for this study (Figure~\ref{MAE_plots}). The numerical values are listed in Tables S20 and S21 in the SI.


The NN model (green in Figure~\ref{MAE_plots}) consistently shows the smallest MAEs for all properties. NN in combination with CM, SOAP, and MIN exhibits similar MAEs for all six properties studied. Only ECFPD leads to slightly larger errors. 

The KRR model (red in Figure~\ref{MAE_plots}) exhibits the second lowest MAE for all properties, except for \tpas{2} (Figure~\ref{MAE_plots}F). In combination with KRR, CM and SOAP show comparable MAE for all properties. 
MIN performs better for \exe{2} and \tpas{2} (panel B and F), and otherwise similarly. 
ECFPD shows the largest errors, except for \exe{2}, where it performs similarly to SOAP (panel B). 

The XGB model (blue in Figure~\ref{MAE_plots}) shows the largest errors for \exe{} and \osc{}. 
Surprisingly, for \tpas{1} it performs similarly to the other models and for \tpas{2} even better than KRR (panels E and F).
The dependence of the errors on the descriptor choice differs from the other models and shows no clear trend.

A comparison of the three general descriptors CM, SOAP, and ECFPD shows the largest average errors in the predictions with ECFPD. However, it is important to consider that ECFPD is based only on the connectivity of the molecules, while CM and SOAP were both derived from the geometry-optimized three-dimensional structures. Thus, ECFPD is significantly less computationally costly, which is an advantage for larger datasets. Therefore, the comparable accuracy of ECFPD relative to the physical descriptors is remarkable. 

Surprisingly, the MIN descriptor performed equally well or better than the other descriptors, although it only has 24 features (1.3\% of CM). 
This shows that bespoke and domain-specific descriptors can be more effective for specific problems. 
However, due to the nature of the descriptor as a collection of the minimal necessary information, it may not be transferable to other data sets. 

To further gauge the importance of the quality of the three-dimensional structure of the molecules in physical descriptors, we also tested predictions using not fully optimized preliminary structures before geometry optimization using DFT (see Dataset Generation).
For CM and SOAP this leads to approximately 20\% larger errors, which in some cases could lead to comparable errors as ECFPD. 
The good accuracy of MIN shows that the use of geometry optimized structures may not be as important as expected.

Another consideration in the choice of the ML models and descriptors is the feasibility in terms of training time and ease of use.
Ranking the ML models in terms of feasibility, we find KRR as the most convenient method, followed by XGB, and lastly NN.
Considering the descriptors, SOAP is significantly more expansive, with about 20 times more features than CM and 8 times more than ECFPD. 
While NN and KRR generally performed better with SOAP, all three models required significantly longer training times compared to the other descriptors.

Considering the absolute errors of the different properties, we see that even the largest MAEs are relatively small compared to the absolute range of the properties (Table S20 and S21, SI). 
In the case of \exe{}, all ML predictions have lower MAE than the error of TDDFT with respect to experimental or high-level quantum chemistry methods, ranging between 0.25--0.3 eV.\cite{Send2011, Jacquemin2009}
For example, the \osc{1}-values range from 0.003--0.6 a.u., while our largest MAE with XBG and ECFPD was only 0.0425~a.u. (7\% of the absolute range).
\tpas{1}-values range between $2$ and $43,000 \;\mathrm{a.u.}$, while our largest MAE with KRR and ECFPD was only 950~a.u. (2\% of the absolute range). 
Thus, the predictions of the photochemical properties are sufficiently accurate for a comparison of candidates. 
Since our objective is to filter out a small number of molecules with maximum values, errors on this scale are insignificant.



\section{Conclusion}
In this study, we have generated and analyzed a comprehensive data set of 2016 potential MNM candidates. We defined key criteria associated with a high photoisomerization quantum yield, leading to unidirectional mechanical motion. These criteria include: (1) achieving high oscillator strength or TPAS for efficient photon harvesting, (2) suppressing back isomerization through well-separated absorption bands to ensure unidirectional rotation, and (3) preserving the $\pi-\pi^*$ -excited state character, as quantified by the PRS, to maximize EZ-isomerization probability.
To this end, we developed the PRS as a metric allowing us to determine whether an EZ-photoisomerization is likely to follow an excitation.

We extensively analyzed relationships between structural features, such as position and chemical nature of the substituents, and the desired target properties for nanomotor applications.

In particular with respect to TPAS, we could identify candidates with significantly (up to two orders of magnitude) increased \tpas{} compared to previously employed molecules in practical two-photon applications.
This shows that substitution with push-pull functional groups is an effective strategy to harness MNMs for two-photon applications. By assessing the PRS, we ensured that substitution does not compromise the photoreactive character of the $\pi - \pi^*$ transition necessary for photoisomerization.

The next step in verifying our proposed strategy, in particular validating the concept of the PRS, is to carry out non-adiabatic ab initio molecular dynamics simulations\cite{Tapavicza2007,Tapavicza2013} with our candidate molecules.  
This will reveal if the PRS is able to reliably predict the desired photoreactivity 
with satisfactory quantum yield, as shown for the unsubstituted motor\cite{lucia2025first}. 
If this is the case, then these structures could be synthesized and tested experimentally. 
In addition, the somewhat arbitrary value of 0.94 that we have chosen for practical purposes as a threshold for the PRS needs to be further validated. 

Furthermore, we showed that ML can be applied to predict the target quantities with excellent accuracy, even with simple descriptors that do not require an optimized three-dimensional structure, as shown by the MIN descriptor. 
This is an important result, in particular for larger chemical spaces, where the computational cost of the geometry optimizations may prevent the screening of candidates.
However, the reason for the good accuracy of the MIN descriptor might be associated with the limited size, homogeneity, and completeness of our dataset. It is unknown if such a minimal descriptor will lead to accurate predictions for larger and more heterogeneous chemical spaces.

Nevertheless, this study serves as a baseline for the screening of candidates for nanomotor applications employing one- and two-photon excitations. 


\newpage



\begin{acknowledgement}

Research reported in this publication was supported by the National Institute of General Medical Sciences of the National Institutes of Health (NIH) under award number R16GM149410. The content is solely the responsibility of the authors and does not necessarily represent the official views of the NIH. We acknowledge technical support from the Division of Information Technology of CSULB.

\end{acknowledgement}

\begin{suppinfo}

Supporting Information with additional results is available in pdf format.

\end{suppinfo}

\section*{Data Availability Statement}
All data, including the Cartesian coordinates of the structures of the molecules, TDDFT photochemical properties, and transition densities given in machine-readable format are available at https://doi.org/10.5281/zenodo.15516776

\bibliography{sources,papers}

\end{document}